
\centerline {\bf Loop Expansion in Light-Cone $\phi^4$ Field Theory}
\vskip 0.2in
\centerline {                   Xiaoming Xu}
\centerline { Institute of Nuclear and Particle Physics, Department of
              Physics, University of Virginia. }
\centerline { Charlottesville, Virginia 22901 }
\centerline { Shanghai Institute of Nuclear Research, Chinese Academy of
              Sciences, Shanghai 201800, China }
\vskip 0.1in
\centerline {                   H. J. Weber   }
\centerline { Institute of Nuclear and Particle Physics,  Department of
              Physics, University of Virginia, }
\centerline {            Charlottesville, Virginia 22901  }
\vskip 0.2in
\centerline {\bf                  Abstract }
\vskip 0.1in
Based on the path integral quantization of the light-cone $\phi^4$ field
theory in 1+1 dimensions, a loop expansion is implemented. The effective
potential as a function of the zero-mode field $\omega$ is calculated up to
two loop order and its derivative with respect to $\omega$ is used to determine
the vacuum expectation value of the field $\phi$. The critical coupling
constant at the occurrence of the spontaneous symmetry breaking is
consistent with that obtained in the ordinary instant-form field theory.
The critical exponents which describe the behavior of the susceptibility and
the vacuum expectation value of $\phi $ near the critical point are evaluated
from the effective potential. The one
loop diagrams for the connected Green's function are calculated in momentum
space. The relevant equal-time correlation function is shown to be closely
related.
\vskip 0.2in
\centerline {\bf     1. Introduction  }
  For many years the light-cone vacuum was considered to be simple: It carries
no longitudinal momentum ($k^+=0$), while Fock operators for fermions have
$k^+ > 0$, and the vacuum state of the free Hamiltonian is an eigenstate of the
light-cone Hamiltonian {\bf with interactions} in many theories. Based on this
physical vacuum a Fock basis can be constructed to produce the hadron
spectrum [1]. However, it was stressed early [2, 3] that zero modes may
change this simple picture drastically, (the zero mode of a field $\phi $ is
defined as $\omega=\lim_{L \to \infty}{1 \over 2L} \int_{-L}^{+L} dx^-
\phi(x)$ to project $\phi $ onto $k^+=0$) and the resulting complexity of
the vacuum has been exploited since then [4-17].
Due to the nontrivial nature of the vacuum the bosonic zero modes have a large
effect on the spectra of the Hamiltonian and field operators such as $\phi$
[13], they enter into internal lines of Feynman diagrams of any order (for
the $\phi^3$ field theory, see [12]), and they
remove certain noncovariant and quadratically divergent terms in the
fermion self-energy in the discretized light-cone quantization of Yukawa
theory [10]; moreover, $\theta$-vacua exist and exhibit non-vanishing
fermion condensates [8, 17].
\par
The quantization of the $\phi^4$ theory has been achieved [3, 4, 8, 13] by
applying the Dirac-Bergmann algorithm [18] for constrained systems.
The vacuum expectation value $<\phi >$ (VEV)
of the scalar field $\phi $ has been calculated by solving
a secondary constraint $\theta_3$ (see Eq. (3)) related to spontaneous
(reflection) symmetry breaking [8, 9, 11, 13].
The constraint equation is most easily obtained from the integration of the
field equation of motion [11].
The vacuum expectation value of the $\phi$ field was obtained in the tree
approximation plus a lowest order correction [8, 11]. In [9] the critical
coupling constant $\lambda_c \approx 40m^2$ is calculated from the curvature
of a potential which has quadratic and quartic terms of the VEV (two
ans\"atze are used for the zero mode field operator). In a series of
studies [13] a so-called $\delta$ expansion combined with numerical
calculations was used to find the VEV as a function of the coupling constant;
a critical coupling constant $\lambda_c \approx 59.5m^2$ was determined
by searching for a converging solution of the constraint equation.
\par
For the $\phi^4$ theory
the $\phi$ field is decomposed into the zero modes and nonzero modes and the
two types of modes couple to each other by a cubic interaction term. This means
they do not move individually but evolve with mutual interactions. Via the
constraint $\theta_3$
the nonzero modes influence the zero modes not only in tree level
interactions but also in higher order diagrams, i.e. loop graphs, and the zero
modes enter into an internal propagator at any order of diagrams describing the
propagation of the nonzero modes [12]. Motivated by this, we first take into
account the vacuum fluctuation based on the classical background field,
i.e. make a loop expansion beyond the tree approximation. The loop diagrams
show a much richer nonlinear structure than the classical approximation
and are responsible for the phase transition between the regime of
spontaneously broken symmetry and its restoration
at finite temperature [19]. Secondly, we calculate the Green's function for
the nonzero modes to one loop order. Following the
Green's function, another important quantity relevant to the nonvanishing
VEV of the field $\phi$, the
equal-time correlation function, is briefly discussed.
This currently interesting physical quantity measures the size of the
domain of disordered chiral condensates formed possibly in ultra-high energy
hadronic or heavy nucleus collisions. In 3+1 dimensions, the zero modes
depend on the other two coordinates $x^1$ and $x^2$. This extra coordinate
dependence substantially increases
the difficulty in evaluating vacuum graphs. Thus for simplicity,
the $\phi^4$ field theory is studied here in 1+1 dimensions with a single
spin-0 field $\phi$.
\par
In [19] a WKB-type expansion is developed for a potential from an effective
action in order to study nonlinear phenomena including spontaneous symmetry
breaking, bound states and resonances, etc. We use this loop expansion to
deal with the nonperturbative effects of $\phi^4$ field theory in 1+1
dimensions. Before starting the loop expansion, the
Lagrangian is shifted by a constant field (in a static case). The propagator
then contains the constant field in its denominator and the coupling
constant in the
cubic interaction depends linearly on the constant field (see, e.g. the third
term in Eq.12). When the constant
field is taken as the vaccum expectation value of a quantum field, the loop
expansion includes the nonperturbative effects of the vacuum represented by
the VEV parameter [20].
\par
This paper is organized as follows. In section 2, we first prove that the
"shift" field technique is applicable in the vacuum persistence amplitude for
the purpose of calculating the VEV. Based on this justification,
we use the loop expansion given by Jackiw [19] and present
the calculation of vacuum diagrams up to two loops. In section 3 the
connected Green's function for nonzero modes at the one-loop order is
calculated in momentum space and the equal-time correlation
function is evaluated for the case of a non-vanishing VEV. In the final section
4, we briefly discuss our results for the critical coupling constant and
critical exponents which describe the behavior of the susceptibility and VEV
near the critical coupling constant.
\vskip 0.2in
\centerline {\bf  2. Effective Potential  }
\vskip 0.1in
The Lagrangian of the light-cone $\phi^4$ field theory in 1+1 dimensions is
$$
{\cal L}(\phi)={1 \over 2} \partial^+\phi \partial^-\phi -{1 \over 2} m^2
 \phi^2  -{\lambda \over 4!}\phi^4              \eqno (1)
$$
where m is the mass and $\lambda$ is the coupling constant.
The field $\phi$ is separated into the zero modes $\omega$ and nonzero modes
$\varphi$: $\varphi=\phi-\omega$, where
$\omega={1 \over 2L} \int_{-L}^{+L} dx^- \phi(x)$ and  L is the boundary of a
spatial box in which the field is quantized. In terms of $\omega$ and
$\varphi$ the Lagrangian is written as
$$
{\cal L}(\varphi, \omega)={1 \over 2}\partial^+\varphi \partial^-\varphi
   -{1 \over 2}m^2(\varphi+\omega)^2 -{\lambda \over 4!} (\varphi+\omega)^4
                        \eqno (2)
$$
Since the light-cone Lagrangian is maximally singular, two primary constraints
are obtained from the definition of the canonical momenta with respect to
$\varphi$
and $\omega$. Applying the Dirac-Bergmann algorithm, three constraints
of second class (denoted by $\theta_1$, $\theta_2$ and $\theta_3$) and their
Dirac brackets were obtained [8]. Alternatively, a symplectic method [21] was
applied to get the same Dirac brackets without the classification of primary
constraints, first and second constraints [22].
The constraint $\theta_3$ is the derivative of the potential with
respect to $\omega$ [9],
$$
\theta_3 = m^2 \omega + {\lambda \over 3!} \omega^3 +{1 \over 2L}
\int_{-L}^{+L} dx^- {\lambda \over 3!} [\varphi^3 (x) +3 \varphi^2 (x) \omega]
\approx 0                       \eqno (3)
$$

{}From these results we write down the vacuum
persistence amplitude in the absence of the external source according to the
formalism of path integral quantization of field theories with second class
constraints [23],
$$
Z(0)=\mid 2Det(\partial^+_x \delta (x-y)) \mid^{1 \over 2} \int {\cal D}\varphi
     exp[{i \over \hbar} \int d^2x {\cal L}(\varphi,\omega)]          \eqno (4)
$$
The determinant before the functional integral is simply a constant, and not
important in calculations of conventional Feynman diagrams. In later
calculations it is neglected. The connected generating functional is
$$
W(0)=-i \hbar~lnZ(0)                  \eqno (5)
$$
and the effective action $\Gamma(\omega)$ is just $W(0)$.
Taking the derivative of $\Gamma(\omega)$ with respective to
$\omega$ gives
$$
{d\Gamma (\omega) \over d\omega} = -{1 \over Z[0]} \int {\cal D}\varphi
exp[-{i \over \hbar} \int dx^+ H_c] \int dx^+ 2L\theta_3,
               \eqno (6)
$$
where $H_c=\int_{-L}^{+L}dx^-[{1 \over 2}m^2(\varphi+\omega)^2
+{\lambda \over 4!} (\varphi+\omega)^4]$ is the canonical Hamiltonian.
It is obvious
that ${d\Gamma(\omega) \over d\omega}=0$ is equivalent to $
\theta_3=0$. The estimate of the vacuum expectation value $<\phi>$
from the constraint $\theta_3=0$ is translated into a calculation of
${d\Gamma(\omega) \over d\omega} =0$. Moreover, it is well-known that the path
integral formalism provides us with a rich knowledge of the field motion and
interactions. We thus expect to obtain more information on the zero-mode
component of the field $\phi$ by carrying out the
loop expansion beyond the classical (tree) approximation. We employ
the frame-independent loop expansion formalism given by Jackiw [19] to get the
VEV and calculate vacuum graphs up to two loop levels.
In Ref. [19], an expansion in powers of
$\hbar$ was made for the effective action $\Gamma(\hat {\phi})$ by shifting
the field $\phi$ by a constant $\hat {\phi}$. In this spirit, Eq. (2)
suggests a "shift" of the field $\varphi$ by a constant field $\omega$.
So in any calculation with the formulas given by Jackiw, it must be kept in
mind that the point $k^+=0$ in the integration over $k^+$ must be left out.
\par
In [12] the secondary constraint
equation for the zero modes and nonzero modes was first solved to express
$\omega$ in terms of $\varphi$, then it was expanded in a power series
of the coupling constant, and finally an expansion in powers of the coupling
constant for the interaction Hamiltonian of the field $\varphi$ was obtained.
In such an expansion the zero modes propagate along
the internal lines in any order of the
Feynman diagrams for the field $\varphi$. In contrast to this procedure, we
first expand the effective action in an $\hbar$ power series, then invoke the
constraint $\theta_3=0$ in this expansion series, or equivalently, implement
${d\Gamma(\omega) \over d\omega}=0$. In other words, according to Eqs.
(3.7)-(3.11) in the second paper of Ref. [8], we first
make an expansion of Z(0) before integrating $\omega$, then carry out the
integration over $\omega$ and thus implement the constraint $\theta_3=0$.
\par
 In the case of an external source J(x), a term $J(x)\varphi(x)$ is
added to the Lagrangian in Eq. (4) to get the vacuum persistence amplitude
Z(J).
In terms of Z(J) the connected generating functional W(J) is defined by
$$
W(J)=-i \hbar~lnZ(J)                \eqno (7)
$$
The effective action is obtained from $W(J)$ by a Legendre transformation
$$
\Gamma(\bar{\varphi})=W(J)-\int d^2 x \bar{\varphi}(x)J(x)        \eqno (8)
$$
with $\bar{\varphi}(x)={\delta W(J) \over \delta J(x)}$. The effective
potential $V(\omega)$ is defined from the effective
action by setting $\bar{\varphi}(x)$ to be a constant field $\omega$, which is
reached by "shifting" the $\varphi (x)$ with the $\omega$ in the Lagrangian,
and extracting an over-all factor of 1+1 dimension volume,
$$
\Gamma (\omega)= -V(\omega) \int d^2 x        \eqno (9)
$$
The ${d\Gamma(\omega) \over {d \omega}}=0$, i.e. ${dV(\omega)
\over d\omega}=0$ produces exactly the VEV of the
field $\phi$. By expanding W(J) and $\delta W(J) \over \delta J$ as a power
series in $\hbar$, the effective potential $V(\omega)$
in a loop expansion is given by [19]
$$
V(\omega)=V_{tree} (\omega) -{1 \over 2} i \hbar \int [d^2k]'~
ln~det~iD^{-1} (\omega;k) +i \hbar <exp[ {i \over \hbar} \int d^2x
{\cal L}_I (\varphi, \omega)]>          \eqno (10)
$$
where $[d^2k]'$ is the measure of the light-cone momentum in 1+1 dimensions
without the point $k^+=0$ since we make the loop exansion for the field
$\varphi$. The first term $V_{tree}(\omega)$ is the classical potential (tree
approximation).
The second term is the one-loop effective potential which involves a logarithm
and the third term
expresses the effective potential generated from multi-loop diagrams.
The ${\cal L}_I(\varphi, \omega)$ is composed of terms cubic and quartic in
$\varphi (x)$.
The $D(\omega;k)$ is the propagator in momentum space and its explicit
dependence on $\omega$ results from the field $\varphi$ "shifted" by
$\omega$. The practical calculation as outlined above starts from the
Lagrangian for the spin-0 $\phi^4$ field,
$$
{\cal L}(\phi)={1 \over 2}\partial^+ \phi \partial^- \phi - {1 \over 2}
(m^2_0 + \delta m^2) \phi^2 - {\lambda_0 + \delta \lambda \over 4!} \phi^4
          \eqno (11)
$$
where the $m_0$ and $\lambda_0$ are the finite but undertermined mass and
coupling constant. The counterterms $\delta m^2$ and $\delta \lambda$ are
given in $\hbar$ power series form. We "shift" the field $\varphi$ by a
constant field $\omega$. The "shifted" Lagrangian is
$$
{\cal L}(\varphi, \omega)={1 \over 2} \partial^+ \varphi  \partial^- \varphi
-{1 \over
2} \mu^2 \varphi^2 -{\lambda_0 +\delta \lambda \over 6} \omega \varphi^3 -
{\lambda_0 + \delta \lambda \over 4!} \varphi^4               \eqno (12)
$$
with
$$
\mu^2=m^2_0 + \delta m^2 + {\lambda_0 + \delta \lambda \over 2} \omega^2
                          \eqno (13)
$$
We see in the shifted Lagrangian the induced mass $\mu$ depends on $\omega
$ and an $\omega$-dependent cubic interaction is obtained. The
$\omega$-dependent  propagator in
the momentum space is defined by the new quadratic term,
$$
D(\omega;p)={i \over {p^+p^- - \mu^2 +i\epsilon}}        \eqno (14)
$$
and the classical potential is
$$
V_{tree}(\omega) ={{m^2_0 + \delta m^2} \over 2} \omega^2
+{{\lambda_0 + \delta \lambda}
\over {4!}}  \omega^4                      \eqno (15)
$$
\par
For a single field $\phi$, the propagator $D(\omega, p)$ is diagonal in
momentum space and the determinant is thus removed.
The one-loop effective potential corresponding to
Fig. 1(a) is written as
$$
\eqalign{
V_1(\omega)
&=-{{i\hbar} \over 2} \int [d^2k]'~ln~iD^{-1}(\omega;k)            \cr
&=- {\hbar \over {8 \pi}} [(2 \Lambda^2 +\mu^2)~ln(2\Lambda^2 + \mu^2)
      -\mu^2~ln\mu^2 ]            \cr}
                            \eqno (16)
$$
where $\Lambda$ is a cut-off of the high momenta $k^+$ and $k^-$.
The two-loop effective potential is
$$
\eqalign{
V_2 (\omega)
&={{\hbar^2 \lambda_0} \over 8} \int [d^2k]'[d^2l]'
   D(\omega; k) D(\omega;l)         \cr
&~~ -{{i\hbar^2 \lambda^2_0} \over 12} \omega^2 \int [d^2k]'[d^2l]'
     D(\omega ; k+l)D(\omega;k)D(\omega; l)   \cr}
                    \eqno (17)
$$
where the first term corresponds to the Fig. 1(b) and the second term the
Fig. 1(c). We obtain
$$
V_2(\omega) = {\hbar^2 \lambda_0 \over 128 \pi^2}~ln^2(1+{2 \Lambda^2 \over
\mu^2}) + {\hbar^2 \lambda^2_0 \omega^2  \over 96 \pi^2} \lbrace ({1 \over
4\Lambda^2} - {\pi \over \mu^2 \sqrt{3+{8\Lambda^2 \over \mu^2} } } )
[ln(1+{2\Lambda^2 \over \mu^2}) + ln(1+{\mu^2 \over 2\Lambda^2})]
+ {2\pi \over 3\sqrt{3}\mu^2}~ln(1+ {\mu^2 \over 2\Lambda^2}) \rbrace
           \eqno (18)
$$
\par
Expanding the counterterms $\delta m^2$ and $\delta \lambda$ in powers of
$\hbar$ gives
$$
\delta m^2 =\hbar \delta m^2_1 + \hbar^2 \delta m^2_2 +\cdot \cdot \cdot ,~~~~~
{}~~~~~\delta \lambda =\hbar \delta \lambda_1 + \hbar^2 \delta \lambda_2 +
\cdot
\cdot \cdot         \eqno (19)
$$
The values of these counterterms are obtained by cancelling divergent terms
appearing in $V_1(\omega)$ and $V_2 (\omega)$ which are expanded to
the order of $\hbar^2$. We find
$$
\delta m_1^2 = {\lambda_0 \over 8\pi}~ln(2e\Lambda^2) + \delta \bar {m}_1^2
                          \eqno (20)
$$
where $\delta \bar{m}_1^2$ is a finite  but arbitrary quantity. This result
resembles other low dimensional calculations, for example, in Ref. [24]. Since
no divergent terms need to be cancelled by $\delta m^2_2$, $\delta \lambda_1$
and $\delta \lambda_2$, we leave them finite but arbitrary. They might be
determined by some renormalization conditions [25] relating to the spin-0 meson
mass and empirical coupling constant. In terms of these quantities, the finite
effective potential up to two-loop orders is
$$
\eqalign{
V(\omega)
&=V_0(\omega) + V_1(\omega) + V_2(\omega)       \cr
&={1 \over 2}m_R^2 \omega^2 +{1 \over 4!} \lambda_R
       \omega^4 + {\hbar \over 8 \pi} a~lna
    +{\hbar^2 \lambda_0 \over 128 \pi^2}~ln^2 a
    + {\hbar^2 \lambda_0 \over 64 \pi^2}~lna +{\hbar^2 \over 8 \pi} (\delta
\bar
    {m}^2_1 + {\delta \lambda_1 \over 2} \omega^2 )~lna      \cr}
             \eqno (21)
$$
with $a=m^2_0 + {\lambda_0 \over 2} \omega^2$, renormalized mass square
$m^2_R=m^2_0 +\hbar \delta \bar {m}^2_1 +\hbar^2 \delta m^2_2$ and
renormalized coupling constant $\lambda_R=\lambda_0 + \hbar \delta \lambda_1 +
\hbar^2 \delta \lambda_2$. The nonlinear logarithmic structure
coming from the loop graphs is explicitly exhibited.
\vskip 0.2in
\centerline {\bf     3. Correlation Function}
\vskip 0.1in
The nonvanishing VEV of the field $\phi$ is related to the
equal-time correlation function. To see this we discuss
the spin-0 particle field with
the momentum $k^+ \geq 0$. The equal-time correlation function in this case is
calculated from
$$
S(x^- - y^-)=<\phi (x) \phi (y)>=<\omega><\omega>
  +<\varphi (x) \varphi (y)>~~~~~~~~~~~~~~~~~~~~~~~~~~~~~~~~~~~~~~~~~~~~~~~
$$
$$
{}~~~~~~~~~=<\phi><\phi> + \lim_{x^+ \to {y^+ + 0^+}} \int_0^{+\infty}
\int_{-\infty}^{+\infty} [d^2p]' [-iG^c_2 (\omega; p)]
e^{-i{p^+ \over 2} (x^- -y^-) -i {p^- \over 2} (x^+ - y^+)}      \eqno (22)
$$
where the $G^c_2(\omega ; p)$ is the connected two-point Green's function in
momentum space. Fig. 2 exhibits three one-loop diagrams which contribute
$$
T_{2a}= - {(\lambda_0 + \delta \lambda)i \hbar \over 8\pi}~ln (1 + {2 \Lambda^2
\over \mu^2}),~~~~~~T_{2b}=0,~~~~~~~
T_{2c}={(\lambda_0+ \delta \lambda)^2 \omega^2 i\hbar \over 16\pi \mu^2
 (1 + {\mu^2 \over 2\Lambda^2}) } \delta_0p^-               \eqno (23)
$$
to the connected Green's function
$$
G^c_2(\omega; p)=D(\omega; p) - (T_1 +T_2 +T_3)D^2 (\omega; p)
                 \eqno (24)
$$
In the expression for $T_{2c}$, the $\delta_{0p^-} = 1$ when $p^-=0$
and vanishes otherwise. This
indicates that the $T_3$ at the momentum
$p^- =0$ gives a nonzero contribution to the connected Green's function at the
order $\hbar$ when the $\Lambda$ goes to infinity. To order $\hbar$, by
cancelling
divergent terms with the counterterms, the connected Green's function is
$$
G^c_2(\omega; p) ={i \over p^+p^--a} +  {i\hbar \over (p^+p^- - a)^2}
[{\lambda_0 \over 8 \pi}~lna
+{\lambda^2_0 \omega^2 \over 16\pi a} \delta_{0p^-}]          \eqno (25)
$$
and the equal-time correlation function is
$$
S(x^- - y^-) = <\phi><\phi> - {i \over 4\pi} \int_0^{+\infty} {dp^+ \over p^+}
e^{-{i \over 2} p^+ (x^- - y^-)}              \eqno (26)
$$
Here the $S(x^- - y^-)$ is time-independent since the field is in equilibrium.
\vskip 0.2in
\centerline {\bf    4. Discussions and Conclusions   }
\vskip 0.1in
Up to this point, we have obtained the effective potential, the connected
two-point Green's function and the equal-time correlation function.
In our calculations, the cut-off
of the $k^+=0$ in the integration over internal momenta does not induce
infrared divergences since the field is massive. The ultraviolet divergence is
cancelled by the counterterms. By solving the ${dV(\omega) \over d\omega} =0$,
the VEV of the field $\phi$ can be obtained.
Numerical calculations are required by the nonlinear formalism
of $V(\omega)$. Moreover, $m^2_0$ may be adjusted in the renormalization
condition [25] so that the constant $a$ in Eq. (21) is always positive to
ensure
that $V(\omega)$ is real. Translation invariance of the theory [3, 12] ensures
that the VEV is a constant. Since the
loop expansion involves integrating over the field $\varphi$, it is somewhat
different from the instant-form field theory. This is a result of the
separation of the zero and nonzero modes and the vacuum structure and implies
that the tadpole diagram is actually excluded, while Figs. 2(a) and 2(c)
remain to contribute to the Green's function, but not to the equal-time
correlation function. The reason is that the equal-time condition
$x^+ - y^+ =0$ and $\delta_{0p^-}$
eliminate their contributions, respectively. The
two-loop order diagrams are expected to contribute a term to Eq. (26),
but are difficult to evaluate. The first term in Eq. (26) is a background
provided by the VEV and is not cancelled by the second term which is produced
by the nonzero modes.
\par
Before the path integral formalism (4) is applied, the limiting
procedure $L \to \infty$ is taken [8]. In this limit the definition for
$\omega$ gives the
L-independent zero-mode field. For a stable field $\phi $, the zero-mode field
$\omega$ is a constant field which is independent of space-time. Notice,
however, that $\omega$ is functionally dependent on the $\varphi$ field through
 the constraint $\theta_3$, and both $\omega$ and $\varphi$ are operators
which include quantum fluctuations. Then
$<\phi>$ is determined from the loop expansion, Eq.10, based on the path
integral formalism while the $\omega$ and $\varphi$ are taken as classical
fields. Nonetheless our result differs from [9]
since the effective potential up to two loop orders
contains the richer nonlinear structure, but is close to a numerical
calculation [13] which is equivalent to using an effective
potential of polynomial form.
\par
The effective potentials in
 the tree approximation, one loop and two loops are plotted
in fig. 3 for a single scalar field $\phi$ with mass $m_{\phi}$ for
which $m^2_R=-m^2_{\phi}$ and $\lambda_R=4m^2_{\phi}$.
We plot the $\omega$ values corresponding
to the minimum of the effective potential which is the VEV
of the $\phi$ field. In the tree approximation the
minimum of the effective potential is about -0.007$GeV^2$ at $\omega \approx
1.2$. The one-loop and two-loop effective potentials give almost the same
minimum value -0.018 at $\omega \approx 1.3$. The one loop
contribution $V_1 (\omega)$ is important while the two loop correction
$V_2 (\omega)$ is small. At $\omega=0$, the three curves reach their local
maximum. If the renormalized mass $m_R=0$,
the maximum and the two minima coincide at $\omega=0$. In this case the
effective potential is a parabola with its minimum located at $\omega=0$, and
$<\phi>=0$. The dash-dotted curve in the one loop approximation almost
coincides with the solid curve in the two loop approximation. This indicates
that the loop expansion series converges. There is also convergence
at the critical point.
\par
The effective potential becomes imaginary at $m_0^2<0$ and infinite
at $m_0^2=0$ when $\omega$ approaches zero. Therefore, we choose $m_0^2>0$.
To start we also assume $\delta \bar {m}^2_1=-(m_0^2+m_{\phi}^2)$,
then we extract the critical coupling constant where the curvature of the
potential at $\omega=0$ changes sign. The second derivative of the
effective potential with respect to  $\omega$ is
$$
{d^2V(\omega) \over d\omega^2}\mid_{\omega=0}
=m^2_R + {\hbar \lambda_0 \over 8\pi}
+({\lambda_0^2 \over 64\pi^2} +{\delta \bar {m}^2_1 \lambda_0 \over 8\pi})
{\hbar^2 \over m_0^2} +{\hbar \lambda_0 +\hbar \delta \lambda_1 \over 8\pi}
lnm_0^2 +{\hbar^2 \lambda^2_0 lnm_0^2 \over 64\pi^2 m_0^2}    \eqno (27)
$$
Setting ${d^2 V(\omega) \over d\omega^2} \mid_{\omega=0} =0$, the critical
coupling constant $\lambda_c$ is
$$
\lambda_c=-{1 \over lnm^2_0}[8\pi m_R^2 +\hbar \lambda_0 +{\hbar^2 \lambda_0
\over m_0^2} ({\lambda_0 \over 8\pi}+\delta \bar {m}_1^2) +{\hbar^2 \lambda^2_0
lnm_0^2 \over 8\pi m_0^2} ]             \eqno (28)
$$
We calculate the critical coupling constant after choosing the values of
$\lambda_0$ and $m_0^2$. The results are listed in Table 1. Clearly,
this effective potential gives critical coupling constants that are
consistent with those in [9, 13] and the values from $22m_{\phi}^2$ to
$55m_{\phi}^2$ reported for the instant-form field theory [26].
\par
The susceptibility above the critical coupling constant is defined as
$$
\chi^{-1}={d^2V \over d\omega^2} \mid_{\omega=0}       \eqno (29)
$$
Near the critical point, the susceptibility and the VEV behave like
$$
\chi^{-1} \propto (\lambda-\lambda_c)^\gamma         \eqno (30)
$$
$$
<\phi> \propto (\lambda_c-\lambda)^\beta          \eqno (31)
$$
The values of the critical exponents $\gamma$ and $\beta$ are shown in Table 1.
If the $\phi^4$ field theory is taken as a representation of the Ising
model in 1+1 dimensions, then $\gamma$ and $\beta$ are 1.75 and 0.125,
respectively [27]. The critical exponents shown in the table are almost
independent of the mass and coupling constant. The critical exponent $\gamma$
is somewhat lower than 1.75. The value of $\beta$ is close to 0.5 obtained in
[13], but much higher than 0.125. We can attribute this discrepancy to the
use of $\delta \bar {m}^2_1=-(m_0^2+m_{\phi}^2)$ and $\delta m_2^2=0$. If we
choose $m_0^2=70m_{\phi}^2$,$\delta \bar {m}^2_1=-50m_{\phi}^2$ and
$\delta m_2^2=-m_0^2-\delta \bar {m}^2_1-m_{\phi}^2$ leading to
$m_R^2=-m_{\phi}^2$, then the critical values become $\lambda_c \approx 30.618
m_{\phi}^2$, $\gamma =1.235$ and $\beta =0.276$, the latter being an
improvement over the 0.5 obtained in mean field theory [27].
\par
In conclusion, we first show the equivalence  between the
${dV(\omega) \over d\omega}=0$ and the constraint $\theta_3=0$; then we
apply the loop
expansion given by Jackiw to calculate the effective potential up to two loop
orders; third, we calculate the equal-time correlation function. The effective
potential is shown to have a nonlinear logarithmic structure. The one loop
contribution $V_1 (\omega)$ is much bigger than the two loop $V_2(\omega)$.
We have applied the effective potential to calculate the critical coupling
constant and two critical exponents which are consistent with other
theories. For a stable field we have given a static description for
spontaneous symmetry breaking in 1+1 dimensions.

\vskip 0.25in
\centerline {\bf          Acknowledgments   }
\vskip 0.15in
X. Xu thanks R. Jackiw, J. Goldstone, Xiangdong Ji, G. Amelino-Camelia and
Chi-Yong Lin for valuable comments and many discussions. We thank Chung-yu Mou
for discussions on the critical exponents. This work is
supported in part by the U.S. National Science Foundation.
\vfill\eject\par

\centerline {\bf                 References }
\vskip 0.15in
\item {[1]}For a review, see: S.~J.~Brodsky and H.-C.~Pauli, Light-cone
quantization of quantum chromodynamics, in Lecture Notes in Physics, Vol. 396
(Springer, Berlin, 1991)
\item {[2]}S.-J.~Chang and S.-K.~Ma, Phys. Rev. {\bf 180}(1969)1506;
T.-M.~Yan, Phys. Rev. {\bf D7}(1973)1780
\item {[3]}T.~Maskawa and K.~Yamawaki, Prog. Theor. Phys. {\bf 56}(1976)270
\item {[4]}R.~S.~Wittman, in Nuclear and Particle Physics on the
 Light Cone, Proceedings of the Workshop, Los Alamos, New Mexico,
1988, edited by M.~B.~Johnson
and L.~S.~Kisslinger (World Scientific, Singapore, 1988)
\item {[5]}A.~Harindranath and J.~P.~Vary, Phys. Rev. {\bf D36}(1987)1141;
{\bf D37}(1988)1076, 3010; C.~J.~Benesh and J.~P.~Vary, Z. Phys.
{\bf C49}(1991)411
\item {[6]}G.~McCartor, Z. Phys. {\bf C52}(1991)611;
           K.~Hornbostel, Phys. Rev. {\bf D45}(1992)3781;~E.~V. Prokhvatilov,
{}~H.~W.~L. Naus and ~H.-J. Pirner, Phys. Rev. {\bf D51}(1995) 2933
\item {[7]}A.~Borderies, P.~Grang$\acute e$ and E.~Werner, Phys. Lett.
{\bf B319}(1993)490; {\bf B345}(1995)458; O.~C.~Jacob, Phys. Lett.
{\bf B324}(1994)149; {\bf B347}(1995)101
\item {[8]}Th.~Heinzl, St.~Krusche and E.~Werner, Phys. Lett. {\bf B256}(1991)
55; {\bf B272}(1991)54; {\bf B275}(1992)410
\item {[9]}Th.~Heinzl, St.~Krusche, S.~Simb\"urger and E.~Werner, Z. Phys.
{\bf C56} (1992)415
\item {[10]}G.~McCartor and D.~G.~Robertson, Z. Phys. {\bf C53}(1992)679
\item {[11]}D.~G.~Robertson, Phys. Rev. {\bf D47}(1993)2549;
\item {[12]}M.~Maeno, Phys. Lett. {\bf B320}(1994)83
\item {[13]}C.~M.~Bender, S.~Pinsky, B.~van~de~Sande, Phys. Rev.
{\bf D48}(1993)816;
S.~Pinsky and B.~van~de~Sande, Phys. Rev. {\bf D49}(1994)2001; S.~Pinsky,
B.~van~de~Sande and J.~R.~Hiller, Phys. Rev. {\bf D51}(1995)726
\item {[14]}A.~C.~Kalloniatis and H.~C.~Pauli, Z. Phys. {\bf C63}(1994)161;
    A.~C.~Kalloniatis and D.~G.~Robertson, Phys. Rev. {\bf D50}(1994)5262;
   R.~W.~Brown, J.~W.~Jun, S.~M.~Shvartsman and C.~C.~Taylor, Phys. Rev.
   {\bf D48}(1993)5873;G.~McCartor and D.~G.~Robertson, Z. Phys. {\bf C62}
   (1994)349
\item {[15]}A.~C.~Kalloniatis, H.-C.~Pauli and S.~Pinsky, Phys. Rev. {\bf D50}
(1994)6633
\item {[16]}K.~G.~Wilson, T.~S.~Walhout, A.~Harindranath, W.-M. Zhang,
R.~J.~Perry and S.~D.~Glazek, Phys. Rev. {\bf D49} (1994)6720
\item {[17]}F.~Lenz, M.~Thies, S.~Levit and K.~Yazaki, Ann. Phys.
{\bf 208}(1991)1
\item {[18]}P.~A.~M.~Dirac, Can. J. Math. {\bf 1}(1950)1; P.~G.~Bergmann, Helv.
Phys. Acta Suppl. {\bf 4}(1956)79; K.~Sundermeyer, Constrained  dynamics,
Lecture Notes in Physics, Vol. 169 (Springer, Berlin, 1982)
\item {[19]}R.~Jackiw, Phys. Rev. {\bf D9}(1974)1686; L.~Dolan and R.~Jackiw,
Phys. Rev. {\bf D9}(1974)3320; G.~Amelino-Camelia and S.-Y. Pi, Phys. Rev.
{\bf D47}(1993)2356
\item {[20]}H.~J.~Schnitzer, Phys. Rev. {\bf D10}(1974)1800, 2042;
L.~F.~Abbott,
J.~S.~Kang and H.~J.~Schnitzer, Phys. Rev. {\bf D13}(1976)2212
\item {[21]}L.~Faddeev and R.~Jackiw, Phys. Rev. Lett. {\bf 60}(1988)1692;
            R.~Jackiw, MIT preprint CTP-2215
\item {[22]}J.~W.~Jun and C.~Jue, Phys. Rev. {\bf D50}(1994)2939
\item {[23]}P.~Senjanovic, Ann. Phys. {\bf 100}(1976)227
\item {[24]}G.~J.~Huish and D.~J.~Toms, Phys. Rev. {\bf D49}(1994)6767
\item {[25]}S.~Coleman and E.~Weinberg, Phys. Rev. {\bf D7}(1973)1888;
P.~M.~Stevenson, Phys. Rev. {\bf D32}(1985)1389
\item {[26]}S.~J.~Chang, Phys. Rev. {\bf D13}(1976)2778; J.~Abad, J.~G.~Esteve
and A.~F.~Pacheco, Phys. Rev. {\bf D32}(1985)2729; M.~Funke, V.~Kaulfass and
H.~Kummel, Phys. Rev. {\bf D35}(1987)621; H.~Kroger, R.~Girard and G.~Dufour,
Phys. Rev. {\bf D35}(1987)3944
\item {[27]}C.~Domb and M.~S.~Green, Phase Transitions and Critical Phenomena,
Vol. 6 (Academic Press, London, 1976); D.~J.~Amit, Field Theory, the
Renormalization Group, and Critical Phenomena, Vol. 6 (McGraw-Hill, New York,
1978); V.~A.~Miransky, Dynamical Symmetry
Breaking in Quantum Field Theory (World Scientific, Singapore, 1993)
\vfill\eject\par
\centerline {\bf    Figure Captions  }
\vskip 0.15in
Fig. 1. (a)one loop. (b)double bubble. (c)"radiatively" corrected single
bubble.
\par
Fig. 2. Feynman diagrams for (a) $T_{2a}$, (b) $T_{2b}$, (c) $T_{2c}$.
\par
Fig. 3. The effective potential as a function of $\omega$ is calculated with
$m^2_0=0.1GeV^2$, $\delta \bar {m}^2_1=-(m^2_0+m^2_{\phi})$, $\delta
{m}^2_2=0$,
$\lambda_0=4m^2_{\phi}$, $\delta \lambda_1=0$ and $\delta \lambda_2=0$ where
$m_{\phi}=m_\pi$ is taken to be the observed pion mass. The dashed
(dash-dotted, solid) curve is the tree (one loop, two loop) approximation.
\vfill\eject\par
\centerline {\bf  Caption for the table}
\vskip 0.15in
Table 1. The $\lambda_0$ and $m_0^2$ are arbitrary but finite quantities.
Results are the critical coupling constant $\lambda_c$ and critical exponents
$\gamma$ and $\beta$.
\vfill\eject\par
$$\vbox{\offinterlineskip
\halign{&\vrule#&\strut\ #\ \cr
\multispan{13}\hfil\bf Table 1\hfil\cr
\noalign{\medskip}
\noalign{\hrule}
height3pt&\omit&&\omit&&\omit&&\omit&&\omit&\cr
&\hfil$\lambda_0$\hfil&&\hfil$m_0^2$\hfil&&\hfil$\lambda_c$
\hfil&&\hfil$\gamma$\hfil&&\hfil$\beta$\hfil&\cr
height3pt&\omit&&\omit&&\omit&&\omit&&\omit&\cr
\noalign{\hrule}
height3pt&\omit&&\omit&&\omit&&\omit&&\omit&\cr
&$45m_{\phi}^2$&&$90m_{\phi}^2$&&$46.29m_{\phi}^2$
&&$1.35$&&$0.676$&\cr
&$45m_{\phi}^2$&&$100m_{\phi}^2$&&$38.55m_{\phi}^2
$&&$1.33$&&$0.675$&\cr
&$30m_{\phi}^2$&&$90m_{\phi}^2$&&$47.42m_{\phi}^2$&
&$1.35$&&$0.676$&\cr
&$60m_{\phi}^2$&&$90m_{\phi}^2$&&$44.58m_{\phi}^2
$&&$1.34$&&$0.675$&\cr
height3pt&\omit&&\omit&&\omit&&\omit&&\omit&\cr
\noalign{\hrule}\noalign{\medskip}
\multispan3\hfil&\multispan{10}\ {~~~~~~~~~~~~~~~~~~~~~~~~~~~~~~~~~~~~~~~
}\hfil\cr}}$$
\bye